\documentclass[superscriptaddress,prl,showpacs,preprintnumbers,amsmath,amssymb,twocolumn,floats]{revtex4}
\usepackage{txfonts}
\usepackage{amssymb}
\usepackage{graphicx}
\usepackage{sidecap}
\usepackage{CJK}
\usepackage{txfonts}

\begin{document}

\title{Observation of Fermi Arcs in non-Centrosymmetric Weyl Semi-metal Candidate NbP}

\author{D. F. Xu$^\sharp$}
\affiliation{State Key Laboratory of Surface Physics, Department of Physics, and Laboratory of Advanced Materials, Fudan University, Shanghai 200433, China}
\affiliation{Collaborative Innovation Center of Advanced Microstructures, Fudan University, Shanghai 200433, China}
\author{Y. P. Du$^\sharp$}
\affiliation{National Laboratory of Solid State Microstructures, Collaborative Innovation Center of Advanced Microstructures, and College of Physics, Nanjing University, Nanjing 210093, China}
\author{Z. Wang}
\author{Y. P. Li}
\affiliation{Department of Physics, Zhejiang University, Hangzhou 310027, China}
\affiliation{State Key Lab of Silicon Materials, Zhejiang University, Hangzhou 310027, China}
\author{X. H. Niu}
\author{Q. Yao}
\affiliation{State Key Laboratory of Surface Physics, Department of Physics, and Laboratory of Advanced Materials, Fudan University, Shanghai 200433, China}
\affiliation{Collaborative Innovation Center of Advanced Microstructures, Fudan University, Shanghai 200433, China}
\author{P. Dudin}
\affiliation{Diamond Light Source, Harwell Science and Innovation Campus, Didcot OX11 0DE, United Kingdom}
\author{Z. -A. Xu}
\affiliation{Department of Physics, Zhejiang University, Hangzhou 310027, China}
\affiliation{State Key Lab of Silicon Materials, Zhejiang University, Hangzhou 310027, China}
\affiliation{Zhejiang California International NanoSystems Institute, Zhejiang University, Hangzhou 310027, China}
\affiliation{Collaborative Innovation Centre of Advanced Microstructures, Nanjing 210093, China}
\author{X. G. Wan}
\email{xgwan@nju.edu.cn}
\affiliation{National Laboratory of Solid State Microstructures, Collaborative Innovation Center of Advanced Microstructures, and College of Physics, Nanjing University, Nanjing 210093, China}
\author{D. L. Feng}
\email{dlfeng@fudan.edu.cn}
\affiliation{State Key Laboratory of Surface Physics, Department of Physics, and Laboratory of Advanced Materials, Fudan University, Shanghai 200433, China}
\affiliation{Collaborative Innovation Center of Advanced Microstructures, Fudan University, Shanghai 200433, China}

\begin{abstract}

We report the surface electronic structure of niobium phosphide NbP single crystal on (001) surface by vacuum ultraviolet angle-resolved photoemission spectroscopy. Combining with our first principle calculations, we identify the existence of the Fermi arcs originated from topological surface states. Furthermore, the surface states exhibit circular dichroism pattern, which may correlate with its non-trivial spin texture. Our results provide critical evidence for the existence of the Weyl Fermions in NbP, which lays the foundation for further investigations.

\end{abstract}


\maketitle

In the past few years, great progress has been witnessed in the study of quantum materials with non-trivial topological electronic structures. By introducing topological orders, insulators can be further classified into trivial one and non-trivial one, that is, topological insulators (TIs) {\cite{TI_1,TI_2}}. Due to their unique physical properties and great potential in applications, TIs have been one of the central research subject of condensed matter physics in the last decade.

The recent proposal that Weyl fermions can be realized in the so-called topological Weyl semi-metals (TWSMs) broadens the classification of topological phases of matter beyond TIs {\cite{TWSM_1,TWSM_2,TWSM_3,R2Ir2O7,HgCr2Se4}}. The low energy excitations in this new type of topological quantum matter are described by the Weyl equation{\cite{Weyl}}. Thus, in the bulk, the conduction and valence bands disperse linearly cross pairs of discrete points (the Weyl points) along all three momentum directions. The Weyl points are associated with a chiral charge that protects gapless surface states (SSs) on the boundary of a bulk sample. These topological SSs take the form of unclosed curves connecting the Weyl points of opposite chirality {\cite{R2Ir2O7}}, leading to the existence of the unique Fermi arcs on the surface. Due to the novel physical phenomena related to the Weyl fermions such as negative magneto-resistivity, quantum anomalous Hall effect and non-local quantum oscillations, the search of TWSMs have attracted worldwide attentions {\cite{WF_Properties_1,WF_Properties_2,WF_Properties_3,WF_Properties_4,WF_Properties_5,WF_Properties_6}}.

\begin{figure}[t!]
\includegraphics[width=8.6cm]{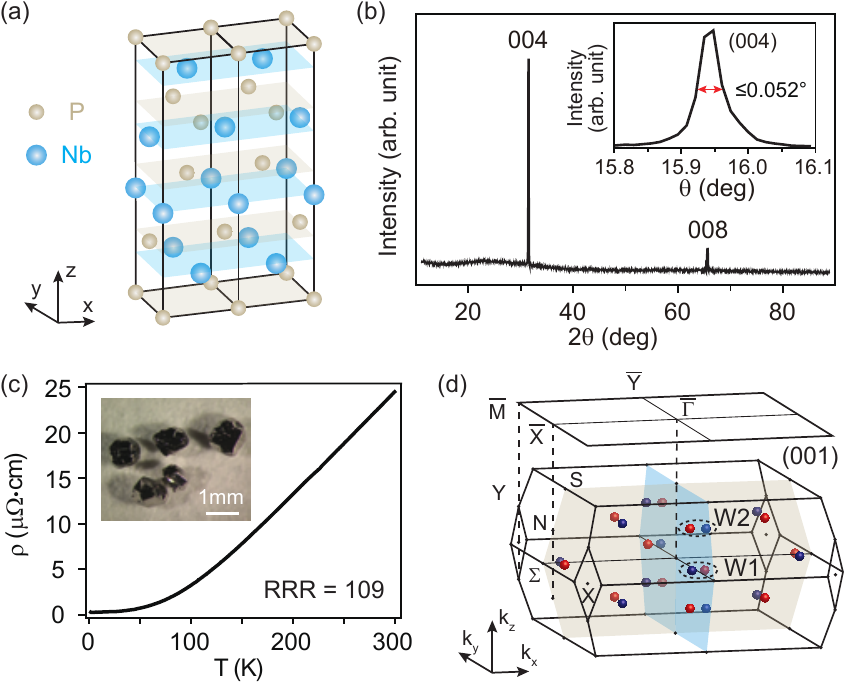}
\caption{(Color online) The crystallographic and transport properties of NbP. (a) Illustration of body-centered-tetragonal structure. (b) The X-ray diffraction pattern of the (001) surface. The full width at half maximum of the (004) peak in the rocking scan is no larger than 0.052$^\circ$. (c) Temperature dependent resistivity curve at zero magnetic field, with RRR over 100. The inset presents the image of the as-grown single crystals, showing flat and shiny surface. (d) Bulk and (001) surface Brillouin zone. Twelve pairs of Weyl points are denoted by red and blue dots to represent opposite chirality. Note, the separation of each pair of Weyl points is exaggerated for clarity.}
\label{lattice}
\end{figure}

\begin{figure*}[t!]
\includegraphics[width=18cm]{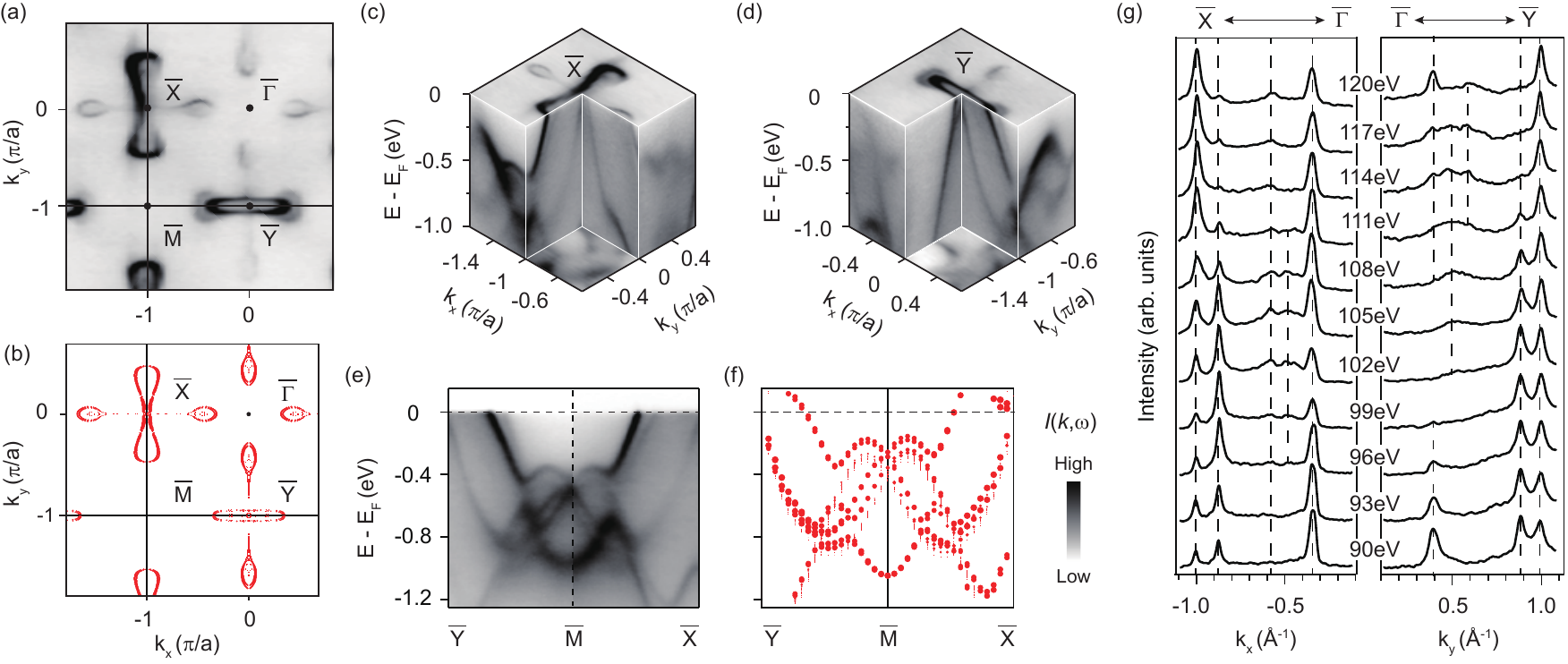}
\caption{(Color online) The measured and calculated surface states (SSs) of NbP on the (001) surface. (a) The photoemission intensity map integrated over [$E_F$ - 10~meV, $E_F$ + 10~meV]. (b) The calculated FS which nicely reproduce the measured one. (c), (d) Three-dimensional intensity plot of the photoemission spectra around the $\overline{X}$ and $\overline{Y}$, respectively. (e), (f) The measured and calculated band dispersions along $\overline{Y}$ - $\overline{M}$ - $\overline{X}$, respectively. The size of the red points in the calculated FS and band dispersions scale with the weight projected on the first unit cell of the P-terminated side. (g) Photon energy dependence of the momentum distribution curves integrated near $E_F$ along high symmetry directions $\overline{\varGamma}$ - $\overline{X}$ and $\overline{\varGamma}$ - $\overline{Y}$, respectively. All the presented data [except those in panel (g)] were taken with circularly-polarized 120~eV photons at 12~K.}
\label{FS}
\end{figure*}

In principle, by breaking either time reversal or spacial inversion symmetry of a Dirac semi-metal (DSM) to remove the spin degeneracy of the bands {\cite{Na3Bi_1,Na3Bi_2,Na3Bi_3,Cd3As2_1,Cd3As2_2,Cd3As2_3,Cd3As2_4,Dirac_Semimetal,3D_Dirac_Semimetal}}, the otherwise degenerate bulk Dirac point will be split into a pair of Weyl points of opposite chirality, leading to the realization of the TWSM. Early predictions of the TWSMs focus on magnetic materials, e.g. R$_2$Ir$_2$O$_7$, HgCr$_2$Se$_4$, which naturally break the time reversal symmetry {\cite{R2Ir2O7,HgCr2Se4}}. However, due to the intricate magnetic domain structure of the samples, the experimental verification of these compounds still remains an open question. Alternatively, the non-centrosymmetric transition metal monoarsenides/phorsphides, TaAs, NbAs, TaP and NbP, have been proposed to be candidates of the TWSMs very recently {\cite{WHM, TaAs_prediction}}. These compounds naturally break the inversion symmetry, meanwhile avoid the complication of working with correlated materials with magnetic ground states, which is ideal for the angle-resolved photoemission spectroscopy (ARPES) measurements. Subsequently, several groups have reported the direct observation of the bulk Weyl points and the surface Fermi arcs by ARPES in TaAs {\cite{TaAs_1,TaAs_2,TaAs_3,TaAs_4}}, NbAs {\cite{NbAs_1}} and TaP {\cite{TaP_1, TaP_2}}, confirming these compounds as TWSMs.  For NbP, although a negative magnetoresistance induced by the chiral anomaly had been observed {\cite{sample_growth}}, an unambiguous evidence about the topological feature is still lack.

On the other hand, NbP exhibit an ultrahigh carrier mobility of 5 $\times$ 10$^{6}$ cm$^{2}$ V$^{-1}$ s$^{-1}$ at 2~K {\cite{NbP_transport,sample_growth}} which is comparable to that of Cd$_3$As$_2$ (9 $\times$ 10$^{6}$ cm$^{2}$ V$^{-1}$ s$^{-1}$ at 5~K) {\cite{Cd_3As_2_mobility}}. Unlike other TDSMs and TWSMs whose carries are either electron or hole, NbP naturally hosts both types of carriers. Consequently, excellent electron-hole compensation leads to the huge magnetoresistance of 850000\% at 1.85~K {\cite{NbP_transport} exceeding that of WTe$_2$ (452700\% at 4.5~K) {\cite{WTe2}. These properties make NbP even more intriguing.

Therefore, probing the SS of NbP is of significant importance not only to verify the topological character of the band structure, but also to understand its novel surface transport. Here, we report the observation of the surface electronic structure of NbP on (001) surface by ultraviolet ARPES. Combining with theoretical calculations, we identify odd number of bands cross the Fermi level in a closed loop, proving the existence of unclosed FS originated from nontrivial surface states. The detailed connectivity of the Fermi arc is given by the calculations, which is qualitatively consistent with our experiment. Furthermore, the SSs exhibit unusual circular dichroism (CD) pattern, which might partially reflected the intrinsic information of the spin texture. Our experimental data and the first principle calculations not only establish that NbP is a TWSM, but also lay foundations for further investigations.

High quality NbP single crystals were synthesized by two-step vapor transport technique as described elsewhere {\cite{sample_growth}}. The crystal structure consists of alternate Nb and P layers, in which the adjacent NbP layers are rotated by 90$^\circ$ and shifted by a/2, leading to a lack of inversion symmetry which is crucial to the realization of the TWSM [Fig.~\ref{lattice}(a)]. Fig.~\ref{lattice}(b) presents the X-ray diffraction (XRD) pattern of our single crystal. The sharp diffraction peaks guarantee the high crystalline quality of the samples. The typical dimension of the single crystal is about 1$mm$ $\times$ 1$mm$ $\times$ 0.5$mm$ [inset in Fig.~\ref{lattice}(c)]. Resistivity data [Fig.~\ref{lattice}(c)] show that NbP is metallic over a wide temperature range (2~K to 300~K), while the residual resistivity ratio [RRR = $\rho$(300~K)/$\rho$(2~K)] is typically about 100. The predicted twelve pairs of Weyl points are illustrated in Fig.~\ref{lattice}(d). For convenience, we denote the Weyl points in the $k_z$ = 0 plane as Weyl point 1 (W1), and the others
are denoted as Weyl point 2 (W2).

\begin{figure*}[t!]
\includegraphics[width=18cm]{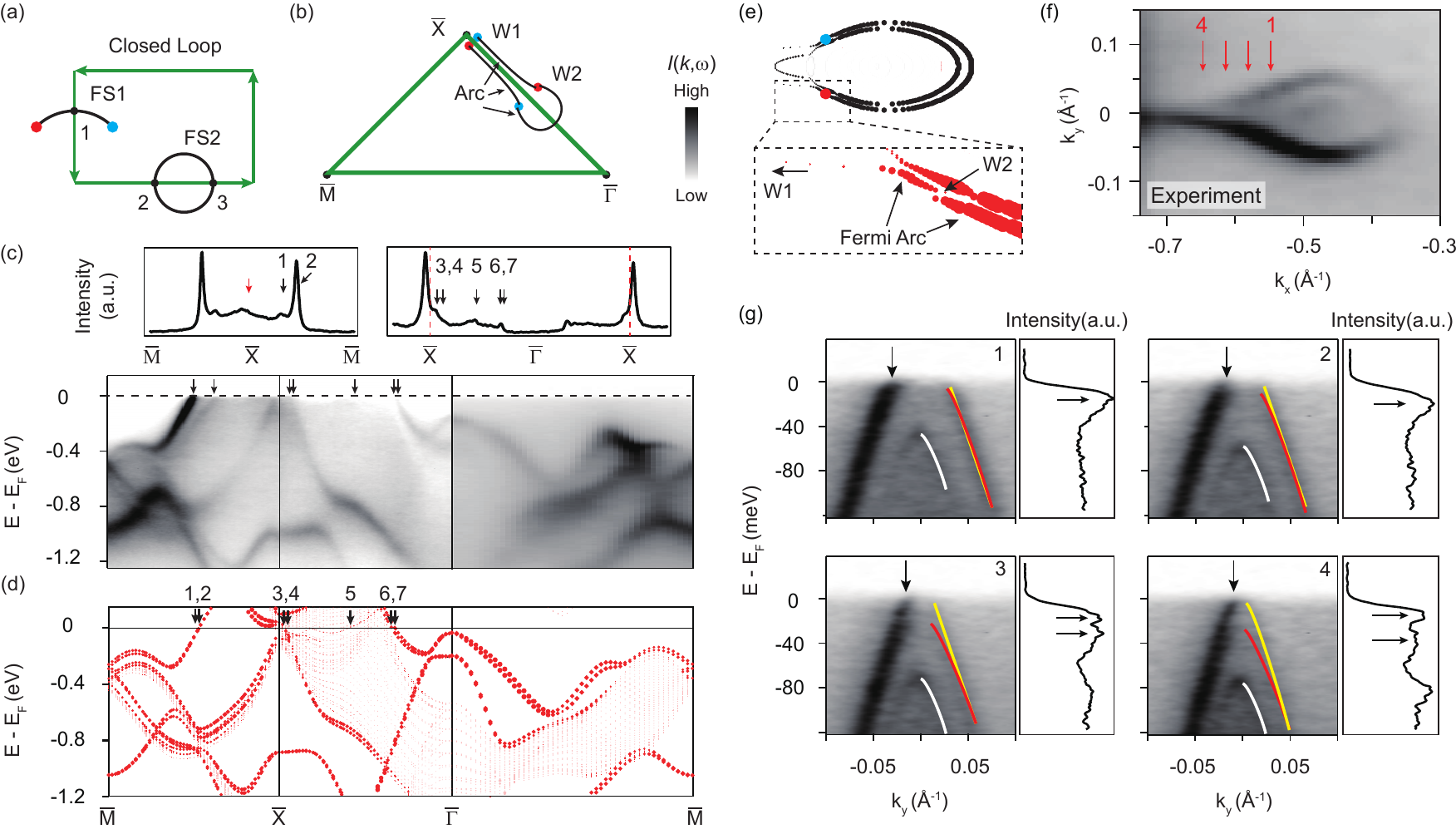}
\caption{(Color online) Observation of the Fermi arc SSs. (a) Schematic of the method to confirm the existence of the Fermi arc. Assuming that the FSs consist of one Fermi arc and one closed pocket (assigned as FS1 and FS2, respectively). Then, choose a closed loop enclosing an odd number of Weyl points (green loop). If there exist a Fermi arc, an odd number of the Fermi crossings in the reference loop (3 times in this case) should be expected. The red and blue points represent the Weyl point with opposite chirality. (b) The illustration of the reference loop $\overline{M}$ - $\overline{X}$ - $\overline{\varGamma}$ - $\overline{M}$ with 3 Weyl points enclosed because the projection of W2 is two-fold degenerate. Note, the separation of Weyl points is exaggerated for clarity. (c) The measured band dispersions along high symmetry directions $\overline{M}$ - $\overline{X}$ - $\overline{\varGamma}$. The upper panel shows the momentum distribution curves (MDCs) integrated near the Fermi level along corresponding high symmetry directions. The peak indicated by the red arrow in the MDCs along  $\overline{M}$ - $\overline{X}$ - $\overline{M}$ corresponds to the residual spectra weight of the bands beneath $E_F$. (d) The calculated band dispersions along high symmetry directions $\overline{M}$ - $\overline{X}$ - $\overline{\varGamma}$, showing excellent correspondence with the experimental data. (e) Calculated FS geometry around the $\overline{X}$. The detailed evolution of different FS segments around the Weyl point is magnified in the inset. The size of the points in the calculated FS and band dispersions scale with the weight projected on the first unit cell of the P-terminated. (f) The measured FS showing overall agreement with the calculations. (g) The measured dispersions along cut 1 to 4 illustrated in panel (f). The corresponding EDCs are integrated around $k_F$ indicated by arrow in each cut. Solid lines are guide of eye for band dispersions.}
\label{FermiArc}
\end{figure*}

ARPES measurements were performed at Beamline I05-ARPES of Diamond Light Source in which VG-Scienta R4000 electron analyzer is equipped. The typical angular resolution is 0.2$^\circ$ and the overall energy resolution is better than 20~meV. Samples were cleaved $in$-$situ$ between adjacent NbP layers along the (001) plane [Fig.~\ref{lattice}(a)], leaving fresh surface which violates $C_4$ rotation symmetry. The measurement pressure was kept below 9$\times$10$^{-11}$ torr.

Our first principle calculation is performed by using the Vienna $ab$ $initio$ simulation package code {\cite{VASP_1, VASP_2}}. An energy cutoff of 400~eV was adopted for the plane-wave expansion of the electronic wave function and the energy convergence criteria was set to 10$^{-5}$~eV. Appropriate k-point meshes of (11 $\times$ 11 $\times$ 1) and (100 $\times$ 100 $\times$ 1) were used for self-consistent and Fermi surface (FS) calculations, respectively. A k-point meshes of (400$\times$400$\times$1) were used to calculate the fine FS around Weyl points. The results are obtained by using the generalized gradient approximation Perdew-Becke-Erzenhof function {\cite{PBE}}. The spin-orbit coupling is taken into account by the second variation method {\cite{SOC}}. To simulate the (001) surface of NbP, we build a slab model composed of a seven unit cell, in which the top and bottom surface are terminated by P and Nb, respectively. A vacuum spacing of 20 $\AA$ is used so that the interaction in the non-periodic directions can be neglected. The experimental lattice parameters were used and the positions of all the atoms were fully optimized. We projected the band structure and FS to the first unit cell of the P-terminated side, in which the size of the red points scales with the projection weight. The obtained results fit the experimental data very well.

The overall photoemission data of NbP are presented in Fig.~\ref{FS}. The measured FS consists of two parts centered around $\overline{X}$ and $\overline{Y}$, respectively [Fig.~\ref{FS}(a)]. In general, each part is composed of two sets of orthogonal paddle-like FSs with some fine features inside, which is nearly in full accordance with the calculations [Fig.~\ref{FS}(b)]. Figs.~\ref{FS}(c) and (d) show the three-dimensional band structure around $\overline{X}$ and $\overline{Y}$, respectively. The measured band dispersions along $\overline{Y}$ - $\overline{M}$ - $\overline{X}$ are again in excellent agreement with the calculations. Here we note that not only the FSs are not four-fold symmetric, but also the band dispersions are anisotropic along $\overline{X}$ - $\overline{M}$ and $\overline{Y}$ - $\overline{M}$. Since the cleaved (001) surface breaks the $C4$ rotational symmetry, these results suggest that the observed band structure may originated from the SSs. The surface origin of the bands can be further confirmed by the photon-energy-dependent ARPES measurements in Fig.~\ref{FS}(g), in which no evident dispersion along $k_z$ can be seen. The intensity variation of some peaks may be due to the matrix element effect. In fact, our photon energy dependence results show that the band structure is dominated by these SSs even in a wider $k_z$ range (90~eV - 220~eV) (not shown here). This may be understood by the intrinsic surface sensitivity of ARPES due to the short escape depth of photoelectrons excited by vacuum ultraviolet photons.

One key signature of a TWSM is the existence of Fermi arcs originated from the topological SSs {\cite{R2Ir2O7}}. To verify its existence, we can choose a closed loop in the surface Brillouin zone with odd number of Weyl points enclosed and count the total time of Fermi crossings. As illustrated in Fig.~\ref{FermiArc}(a), the FS can only cross the loop an even number of times unless it is unclosed (Fermi arc). Although theoretically feasible, it is challenging to experimentally obtain the correct number of crossings in TaAs class of materials due to the extreme proximity of the W1 points and the degeneracy of the bands {\cite{TaP_2}}. To settle this problem, experimental data are commonly combined with calculations to help to precisely identify the Fermi crossings as exemplified in ref \cite{TaAs_2} and \cite{TaAs_4}. In the present case, we choose $\overline{M}$ - $\overline{X}$ - $\overline{\varGamma}$ - $\overline{M}$ as reference loop [Fig.~\ref{FermiArc}(b)] with 3 Weyl points included. The measured band dispersions are presented in Fig.~\ref{FermiArc}(c), which is in excellent correspondence with the calculations [Fig.~\ref{FermiArc}(d)]. We can identify 2, 5 and 0 crossings with the assistance of our calculations along $\overline{M}$ - $\overline{X}$, $\overline{X}$ - $\overline{\varGamma}$ and $\overline{\varGamma}$ - $\overline{M}$, respectively. Thus, the surface FSs cross the reference loop 7 times, providing direct evidence of the existence of Fermi arcs on the (001) surface of NbP.

\begin{figure}[t!]
\includegraphics[width=8.6cm]{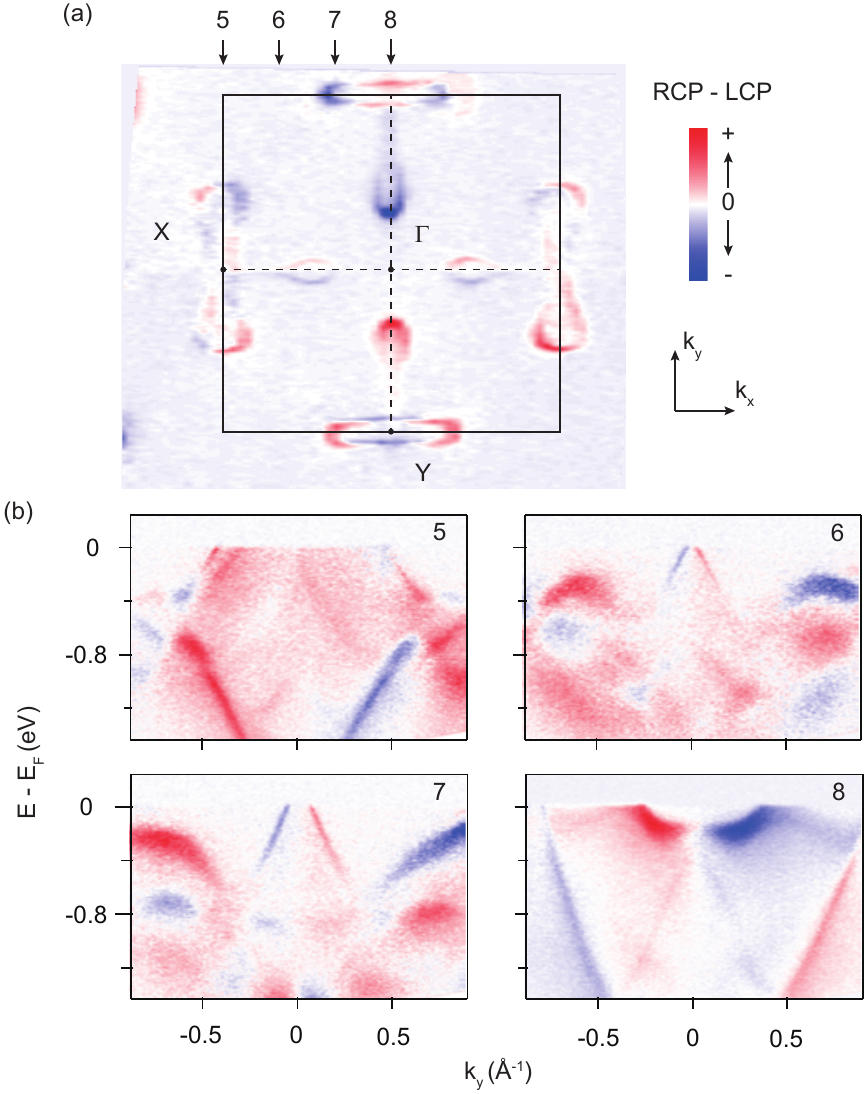}
\caption{(Color online) Circular dichroism of the electronic structure. (a) The Differential map of the RCP and LCP photoemission intensities taken with 90~eV photons. (b) The corresponding intensity plot along several cuts indicated in panel (a).}
\label{CD}
\end{figure}

Compared with the separation of the W1 points, that of the W2 points are larger. To examine the detail connectivity of the Fermi arc, we zoom in around the W2 Weyl point to check the SSs in higher resolution [Figs.~\ref{FermiArc} (e) - (g)]. In Fig.~\ref{FermiArc}(e), our calculations show that the paddle-like FSs are actually composed of two dominant oval FSs. The inner one forms closed FS contributed by trivial SS. The outer one consists of three segments (Fermi arcs), two of which connect the W1 and W2 points, while the rest one directly connects the W2 points [Fig.~\ref{FermiArc} (b)]. The Fermi arc connectivity here resembles that observed in NbAs {\cite{NbAs_1}}. Limited by the experimental resolution, the measured FS only exhibits one broad oval feature [Fig.~\ref{FermiArc}(f)]. To prove the correspondence with calculations, in Fig.~\ref{FermiArc}(g), we check the band evolution around the neck of the paddle-like FSs. From cut 1 to 4 [Fig.~\ref{FermiArc}(g)], a barely-resolved band (in red) can gradually be separated from the outer one (in yellow). This can be further verified by the energy distribution curves (EDCs) integrated around $k_F$ in each cut. The broad EDCs peak in cut 1 and 2 spilt into 2 peaks in cut 3 and 4, suggesting that the broad feature is indeed contributed by two bands. The inner band further sinks below $E_F$ forming closed FS, which is consistent with our calculations. Meanwhile, the outer band further extends toward the W1 points. Although an open Fermi arc is not observed experimentally as them connects W1 and W2, the excellent correspondence between experiment and calculations establishes the existence of Fermi arc SSs in NbP.

In TaAs class of materials which lack inversion symmetry, calculations show that Rashba-type spin polarization naturally exists in the the SSs and leads to strong spin textures {\cite{TaAs_4}}. With the presence of spin-orbit coupling, both spin and orbital angular momentum (OAM) of a state would exhibit conjugate textures around the FS. To further explore the spin texture of the SSs, we performed CD-APRES experiment which has been demonstrated to be a powerful tool to probe the OAM texture of the surface states in TIs {\cite{CD_1,CD_2}}.

Fig.~\ref{CD}(a) show the FS mapping of the difference under the right-circular polarized (RCP) and left-circular polarized (LCP) light. The intensity map presents a strong difference between the RCP and the LCP data, which can be further verified by several photoemission intensity plots along cut 5 to 8. Around the Y point, the intensity of the oval FS sudden change its sign. This suggests that the CD observed here is most likely due to different OAM, rather than some experimental geometry effects. Overall, the CD pattern is antisymmetric with respect to $\overline{\varGamma}$, which can be explained by the time reversal symmetry. Due to the enormous amount of calculation needed, we leave the quantitative analysis of the CD to the future.

To summarize, we have observed SSs on the (001) surface of NbP by using ultraviolet ARPES. Combined with calculations, we identify the existence of the Fermi arc originated from topological SSs. Furthermore, we present the CD of the SSs, which may reflect the intrinsic information about the spin texture. Our results not only provide critical evidence to prove that NbP is a TWSM, but also lay foundation for further investigations.

$^\sharp$ These authors contributed equally to this work.

We thank Diamond Light Source for access to Beamline I05-ARPES (Proposal No. SI11914) that contributed to the results presented here.

This work is supported by National Natural Science Foundation of China under Grant Nos 11174124, 11274068, 11374137, 11421404 and 13ZR1451700, and the National Basic Research Program of China (973 Program) under the grant No. 2012CB921402.

\end{document}